\documentclass[journal]{vgtc}                     % final (journal style)
\onlineid{1773}

\vgtccategory{Research}

\vgtcpapertype{Application/Design Study}

\title{\system{}: A Visual Approach for Building, Managing, and Analyzing Weather Simulation Ensembles at Runtime}

\author{%
Carolina Veiga Ferreira de Souza, Suzanna Maria Bonnet,\\
Daniel de Oliveira, Marcio Cataldi, Fabio Miranda, and Marcos Lage
}

\authorfooter{
  \item de Souza is with the Universidade Federal Fluminense and the University of Illinois. E-mails: carolinavfs@id.uff.br, carolvfs@illinois.edu.
  \item Bonnet is with the Universidade Federal do Rio de Janeiro.
  \item de Oliveira, Cataldi, and Lage are with the Universidade Federal Fluminense. E-mail: mlage@ic.uff.br.
  \item Fabio Miranda is with the University of Illinois Chicago.
}

\abstract{
Weather forecasting is essential for decision-making and is usually performed using numerical modeling. Numerical weather models, in turn, are complex tools that require specialized training and laborious setup and are challenging even for weather experts. Moreover, weather simulations are data-intensive computations and may take hours to days to complete. When the simulation is finished, the experts face challenges analyzing its outputs, a large mass of spatiotemporal and multivariate data. From the simulation setup to the analysis of results, working with weather simulations involves several manual and error-prone steps. The complexity of the problem increases exponentially when the experts must deal with ensembles of simulations, a frequent task in their daily duties. To tackle these challenges, we propose ProWis: an interactive and provenance-oriented system to help weather experts build, manage, and analyze simulation ensembles at runtime. Our system follows a human-in-the-loop approach to enable the exploration of multiple atmospheric variables and weather scenarios. ProWis was built in close collaboration with weather experts, and we demonstrate its effectiveness by presenting two case studies of rainfall events in Brazil.
}


\teaser{
  \centering
  \includegraphics[width=\linewidth]{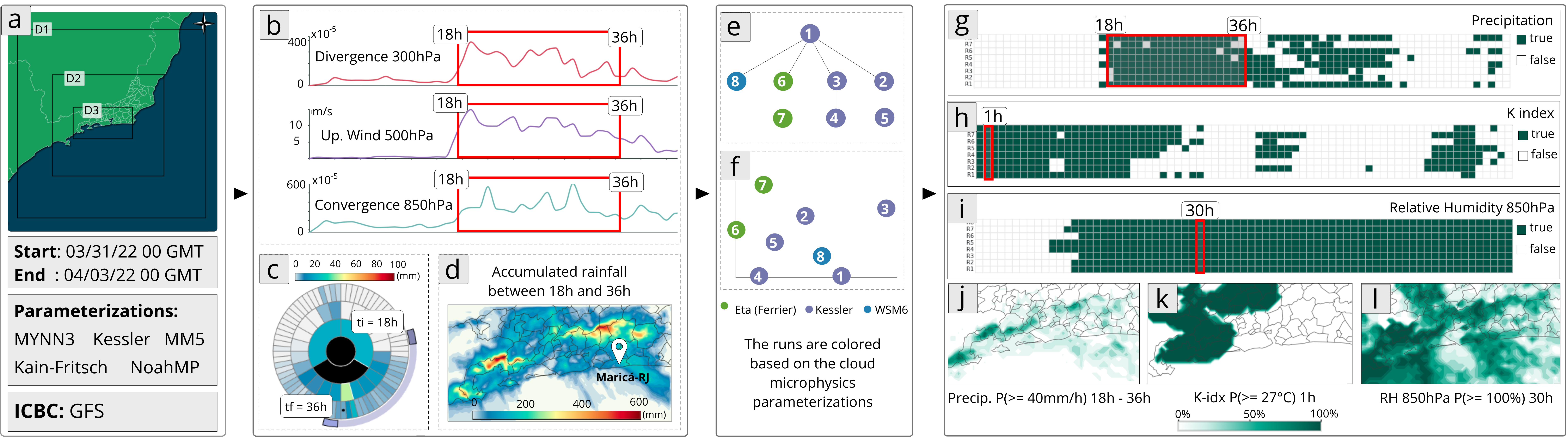}
  \caption{Building, managing, and analyzing a \textcolor{black}{Weather Research and Forecasting (WRF)} model ensemble using \system to study an extreme rainfall event in Maricá (RJ, Brazil) in April 2022. Before running the first simulation, weather experts defined the simulation domain, a 72-hour time horizon, the initial and boundary conditions data, and a set of physical parameterizations~(a). After running the first simulation, using the line charts panel, they identified relevant predictions of divergence at 300~hPa, convergence at 850~hPa, and vertical upwind at 500~hPa between 18h and 36h~(b). By selecting this time interval in the sunburst chart~(c), they saw that the model predicted heavy rain for several locations but underestimated the volume in the city (white marker in~(d)). Using the runs overview graph~(e), the experts generated a simulation ensemble for a broader study. No pattern among the members was observed in the rain volume results~(f). The experts then inspected the heat matrices to verify the chances of having a minimum rainfall of 40~mm/h between 18h and 36h~(g). However, in this period, the probability of this scenario was small in Maricá~(j). Similarly, a minimum K-index of 27°C until 27h~(h) and relative humidity of 100\%~(i), two conditions related to extreme rainfall, could be observed only outside the city~(k,~l). The results show that the model underestimated the event's disaster potential in Maricá.
}
\label{fig:teaser}
}

\graphicspath{{figs/}{figures/}{pictures/}{images/}{./}} % where to search for the images

\usepackage{tabu}                      % only used for the table example
\usepackage{booktabs}                  % only used for the table example
\usepackage{lipsum}                    % used to generate placeholder text
\usepackage{mwe}                       % used to generate placeholder figures
\usepackage{mathptmx}                  % use matching math font

\usepackage[T1]{fontenc}    
\usepackage[utf8]{inputenc}
\usepackage{xspace}
\usepackage{mathtools}
\usepackage[normalem]{ulem}

\newcommand{\hidecomment}[1]{}
\newcommand{\system}{\textsc{ProWis}\xspace}

\newcommand{\myparagraph}[1]{\vspace{0.05cm}\noindent \textbf{#1}\xspace}

\begin{document}

%%%%%%%%%%%%%%%%%%%%%%%%%%%%%%%%%%%%%%%%%%%%%%%%%%%%%%%%%%%%%%%%
%%%%%%%%%%%%%%%%%%%%%% START OF THE PAPER %%%%%%%%%%%%%%%%%%%%%%
%%%%%%%%%%%%%%%%%%%%%%%%%%%%%%%%%%%%%%%%%%%%%%%%%%%%%%%%%%%%%%%%

\firstsection{Introduction}
\maketitle

%%%%%%% Paragraph
Weather conditions significantly impact agriculture, transportation, public safety (\textit{e.g.,} during extreme climate events), and many other critical areas, making forecasting indispensable to designing operational strategies, enabling weather-resilient services, helping decision-making, and defining public policies~\cite{wmo1999, mizutori2020}. 
Although accurate weather forecasting is crucial, it is still a challenging and active research topic. To perform weather predictions, climate specialists must consider an enormous amount of data (\textit{e.g.,} from satellites, weather stations), construct ensembles of numerical simulations, and rely on past experience. 

%%%%%%% Paragraph
Ensembles of numerical weather simulations are computed using mathematical models that describe atmospheric behavior through equations based on physical laws~\cite{warner2010}. These models depend on initial conditions, terrain characterization, parameterization of physical processes, and discretization strategies. 
Weather experts must go through dependent and complex steps to run a simulation. These steps and their data dependencies resemble large-scale scientific workflows~\cite{oliveira2019}.
First, meteorologists must define the simulation domain considering terrain characteristics, land-use information, and spatial discretization aspects. Then, they configure the time horizon and discretization to be used during the simulation. Based on the spatiotemporal setup, the user must set the initial and boundary atmospheric conditions. Finally, previous steps are considered to specify the dynamic/thermodynamic behaviors and the micro-scale processes.

Usually, setting up these steps requires editing large text files to define several parameters. Also, to run each step, the weather expert must execute various commands in a terminal environment. 
Orchestrating the setup and running of these steps can become laborious and error-prone if manually performed. For example, the weather forecasting community has widely adopted an important class of models, known as limited-area models, since it allows running simulations at higher resolutions than global models~\cite{davies2014}.
However, configuring the simulation domain demands the conciliation of technical nuances like the selection of projection methods, the verification of nested grid coherence, \textit{etc.}

Even though some open and commercial software provide tools~\cite{wizard} to facilitate the configuration and execution of a simulation, they are still complex to use, especially when the researcher's goal is to build an ensemble of simulations considering multiple configurations. In this scenario,  members of an ensemble may use shared configurations and steps, which opens opportunities for developing solutions that take better advantage of the available computational resources.
However, running a single weather forecast is already time-consuming, let alone building and managing simulation ensembles.
The ability to orchestrate multiple runs simultaneously, easily start new simulations, inspect partial forecast results at runtime, and cancel unpromising runs may enable faster and more precise analyses.
Moreover, analyzing the output of weather simulations requires the inspection of multiple variables (rain volume, temperature, \textit{etc.}) over space and time. This task becomes more arduous when a simulation ensemble is considered since they allow the study of uncertainties in the outputs. To do so, it is necessary to compute statistics and probabilistic information of groups of ensemble members on demand, which is also challenging to do manually.

%%%%%%% Paragraph
Given the aforementioned challenges involved in building, managing, and analyzing ensembles of weather simulations, new approaches to make the work of weather forecasters less manual, time-consuming, and error-prone may significantly impact their workflow.
Also, new approaches can potentially increase meteorologists' ability to interpret models, calculate risks, and identify relevant weather scenarios.

%%%%%%% Paragraph
In this work, we studied the analysis processes of weather experts and how to combine knowledge from meteorology, atmospheric modeling, visualization, and provenance to help make weather predictions more efficient and reliable.
From that, we built \system, a visual analytics system to assist weather professionals to work with the Weather Research and Forecasting model (WRF)~\cite{grell2005}. Our contributions~are:
\begin{itemize}[noitemsep,topsep=0pt,leftmargin=*]
    \item A collection of visual elements that allow specialists to easily configure and run WRF simulations by setting parameters such as time horizon, simulation domain, initial/boundary conditions, and micro-scale physical phenomena representation approaches.
    \item The use of workflows to facilitate the construction of simulation ensembles. Such workflows are managed by a Workflow Management System (WfMS), which captures and stores provenance data (\textit{i.e.}, the data derivation process~\cite{koop2018}) to allow for integrated analysis and reproducibility of simulation results. 
    \item A set of interactive visualizations that enable the exploration of multiple atmospheric variables and the investigation of weather scenarios at runtime. The visualizations can adapt to represent single or multiple ensemble members.
    \item Two case studies performed in collaboration with weather experts demonstrating the effectiveness of \system.
\end{itemize}
In summary, this paper presents \system, a web-based system that allows the setup, execution, management, tracking, and exploration of simulation ensembles through a visual, interactive, and integrated evaluation of the multivariate and spatiotemporal output data at runtime.

\section{Related work}
\label{sec:related}
This work contributes to several aspects of the weather forecasting process: setup of WRF simulations, orchestration of runs, monitoring of simulations at runtime, and visualization of simulation ensembles.

To the best of our knowledge, the primary tool to aid WRF simulations is the so-called WRF Domain Wizard~\cite{wizard}, a graphical interface that assists in the definition of any WRF parameter and can run individual  WRF simulations. 
However, since it is a general-purpose tool, it is still a laborious and complex task to build and run multiple simulations (a common scenario in the daily duties of weather professionals) using the tool.  
For example, researchers working in specific regions, \textit{e.g.}, Brazil, rarely need to change the geographic projection method used by the model. More generally, it is reasonable to expect some model settings to be preserved throughout several runs, assuming default values. These usage features are not present in the WRF Domain Wizard, which requires the definition of all simulation parameters at every new simulation setup. 
In this sense, user-friendly WRF configuration tools that facilitate the study of scenarios and enable the management, monitoring, and analysis of multiple simulations at runtime can help experts save time and not expose the simulations to human-made errors. Previous commercial or academic tools do not support these features.

Atmospheric sciences are closely related to visualization since the latter effectively bridges the gap between simulation data and knowledge about climate conditions.
Rautenhaus \textit{et al.}~\cite{rautenhaus2018} reported the need to bring together specific concepts of meteorology and state-of-the-art visualizations to enhance the analytical skills of weather professionals. 
However, one of the challenges visualization researchers face is that experts in this domain resist adopting new interactive visualizations. Visualization researchers must be aware of the demands and fears of domain experts so they can focus their efforts on increasing the acceptance of visualization tools that enable superior analysis capabilities. 
Most tools commonly used in meteorological data exploration are command-line-based pieces of software that implement functions for data import, statistical analysis, and image generation (\textit{e.g.}, Ferret~\cite{ferret}, GrADS~\cite{grads}, GMT~\cite{gmt}). 
Recently, Nikfal~\cite{nikfal2023} proposed a tool called PostWRF aimed at helping WRF users visually handle and explore outputs. Even though PostWRF facilitates the work of some experts, it can be considered a technical tool since it requires programming skills.

\myparagraph{Ensemble visualization.} Several works contributed to the visualization of weather simulation ensembles.
Potter \textit{et al.}~\cite{potter2009} developed Ensemble-Vis, a visualization system for analyzing climate ensembles considering different initial conditions. It was one of the pioneers in showing the utility of interactive and linked structures for exploring the average behavior of multiple atmospheric fields in space and time.
Sanyal \textit{et al.}~\cite{sanyal2010} proposed a tool that exploits the advantages of visual structures such as glyphs and ribbons for uncertainty studies of ensembles. 
Cox \textit{et al.}~\cite{cox2013} presented an alternative visualization technique to support the forecast of hurricanes. They primarily used isocontours, which appear and disappear dynamically as the risk varies.
\textcolor{black}{Waser~\textit{et al.}~\cite{waser2011} presented a data-flow-based approach for studying uncertainty and constructing simulation ensembles.}
Diehl \textit{et al.}~\cite{diehl2015} created an interactive visual dashboard to help experts that use WRF to study its outputs. It uses small maps strategically placed on timelines to provide an overview of model outputs and enable the identification of visual patterns.
Rautenhaus \textit{et al.}~\cite{rautenhaus2015} proposed an open-source application that provides statistical and probabilistic ensemble analysis based on 2D and 3D visualizations.
Biswas \textit{et al.}~\cite{biswas2017} proposed an interactive visual system that allows the analysis of multiple ensembles.
Wang \textit{et al.}~\cite{wang2017} developed a visual strategy to explore correlations among simulation parameters.
Watanabe \textit{et al.}~\cite{watanabe2022} proposed an angle-based parallel coordinate graph for exploring large sets of simulations. 
de Souza \textit{et al.}~\cite{souza2022} developed a visual system that enables the analysis of large ensembles of extreme event simulations.

\myparagraph{Provenance-aided visualization.} Previous research achieved positive results by combining data visualization and provenance capabilities, including the visualization of weather and climate data~\cite{santos2013, williams2013,behrens2018}).
Callahan \textit{et al.}~\cite{callahan2006} proposed VisTrails, a workflow management system that orchestrates the execution of multiple tasks and captures provenance data to support the exploration and comparison of data, simulations, visualizations, \textit{etc.} 
Santos \textit{et al.}~\cite{santos2013} used provenance features from VisTrails to support the visual exploration of climate data.
Stitz \textit{et al.}~\cite{stitz2019} combined data visualization and provenance control to record interactions, restore previously accessed views, and find similar analyses.
Gratzl \textit{et al.}~\cite{gratzl2016} introduced a model that combines  data exploration and presentation through visual stories generated by the user's analysis history and interactivity.
Xu \textit{et al.}~\cite{xu2018} created a system to assist team analysis of complex data exploration. 
Behrens \textit{et al.}~\cite{behrens2018} presented a system aimed at managing, creating, and maintaining sets of simulations of coupled models. Their work used provenance control to prevent repetitive steps throughout the analyses.

Although some previous research related to our work exists, none addresses our main contribution: the proposal of a visualization and user-friendly system designed to help weather experts set up, execute, manage, track, and visually explore the simulation ensembles at runtime. 
As shown in the case studies (Section ~\ref{sec:case}), \system allows professionals to perform detailed analyses that would be difficult to achieve without the tool. 

\section{Background}
\label{sec:background}
%
%%%%%%%%%%%%%%%%%%%
% NUMERICAL MODELS
%%%%%%%%%%%%%%%%%%%
%
\myparagraph{Numerical models.} 
A Numerical Weather Model (NWM) can approximate the solution of complex mathematical equations that describe physical behaviors such as thermodynamic laws, Newton's second law, and the continuity equation, which can depict the atmosphere's state at a particular time given the initial and boundary conditions.  
To do so, the area of interest must be represented as a three-dimensional grid used to discretize the equations. The grid's resolution influences the quality of the solution, but using high resolutions may be computationally unfeasible if considering large areas. Limited-area NWMs (NWMs considering restricted geographic areas, small-scale phenomena, and short periods, such as days or hours) are widely adopted to overcome this limitation. The initial conditions needed to solve limited-area NWMs comprise samples of the atmospheric state, measured \textit{in-situ} by meteorological instruments and data extracted from satellite data. Also, boundary conditions are obtained from global NWMs, \textit{i.e.,} NWMs that consider the entire globe as the simulation domain but computed using coarser grids. They are constantly updated by global organizations such as the National Centers for Environmental Prediction~\footnote{https://www.ncdc.noaa.gov/} and the European Centre for Medium-Range Weather Forecasts~\footnote{https://www.ecmwf.int/}.
When the spatial resolution used to generate the initial and boundary conditions is much coarser than the resolution of interest in the limited-area NWM, using nested domains to smooth the data is common. Suppose the global NWM utilizes a resolution of $25~km$, and the expert wants to build a grid of $2~km$ in the limited-area NWM. In that case, it is possible to set up nested grids of $18~km$, $6~km$, and $2~km$ and solve the model at multiple resolutions, using the solution of the coarser resolutions as the initial/boundary condition of the finer ones.

\begin{figure}[!ht]
\centering
\includegraphics[width=.85\linewidth]{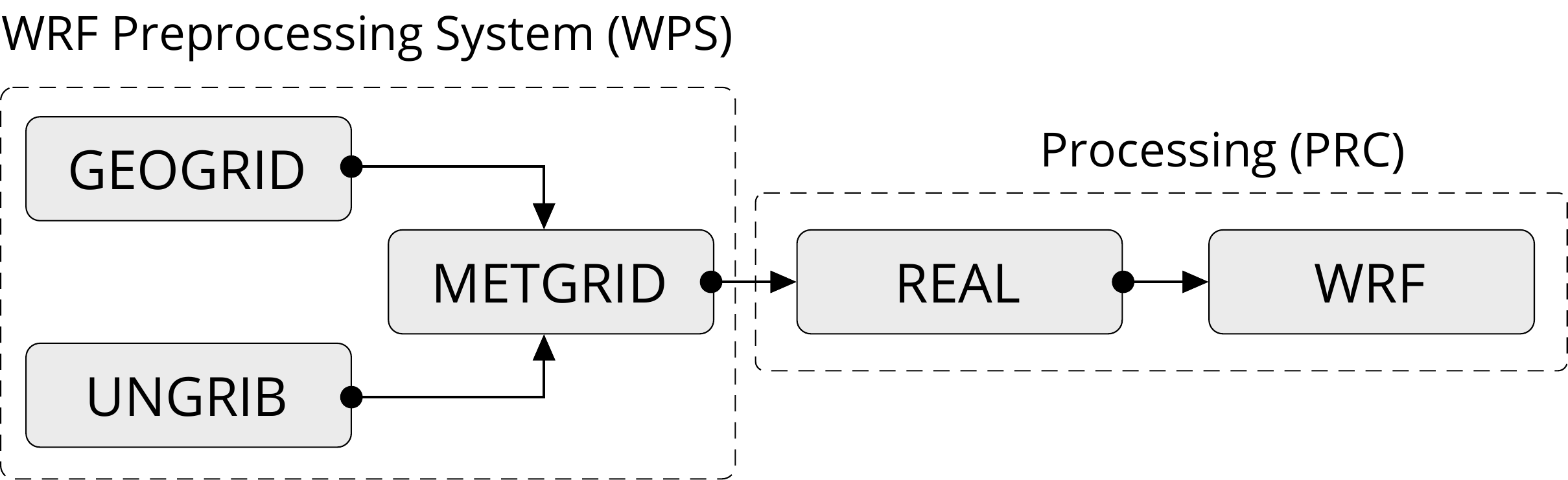}
\caption{WPS and PRC sub-workflows. Geogrid, Ungrib, and Metgrid process terrain, ICBC, and meteorological data. Real and WRF consume the WPS output and perform the simulation.}
\vspace{-0.2cm}
\label{fig:wpswrf}
\end{figure}

%%%%%%%%%%%%%%%%%%%
% WRF
%%%%%%%%%%%%%%%%%%%
%
\myparagraph{Weather Research and Forecasting workflow}.
The WRF model~\cite{skamarock2005} is a numerical weather model widely adopted for research and operational purposes by several areas that depend on the weather, such as for atmospheric chemistry~\cite{grell2005}, hydrological modeling~\cite{gochis2020}, and wildland fires~\cite{coen2013}. WRF supports limited-area NWM simulations and provides a free and open-source implementation that can be integrated into other platforms. 
To run a WRF simulation, a well-defined workflow must be executed. This workflow, depicted in~\autoref{fig:wpswrf}, can be decomposed into two connected sub-workflows named WRF Preprocessing System (WPS) and Processing (PRC). Each sub-workflow is, in turn, composed of multiple steps and their data dependencies. 
Running the complete workflow requires editing several configuration files, downloading initial and boundary conditions (ICBC) data, and executing multiple WPS and PRC programs using the command line.
Most of the WPS and PRC configuration files' content is organized by columns, each representing one nested domain. Using these files, users can configure different simulations on each nested domain. 
WPS sub-workflow consists of three steps, each one associated with the execution of a program: (a)~\textit{Geogrid} defines the domain(s) discretization and interpolates static data, such as topography and land use categories, over the domain; (b)~\textit{Ungrib} processes the ICBC data; (c)~\textit{Metgrid} receives the Geogrid/Ungrib outputs, and horizontally interpolates the meteorological information over the domain(s). 
The PRC sub-workflow is composed of two steps: (a) \textit{Real} program receives the Metgrid's outputs and vertically interpolates meteorological fields, and (b) the \textit{WRF} implementation uses the output of the Real program to start a simulation. The simulation execution time depends on the number of domains, the spatiotemporal discretization, and the computational environment, among other variables. 
The final simulation output is a collection of files in the NetCDF~\cite{netcfd} format. Each file corresponds to the simulation output obtained using one domain. Three NetCDF output files will be produced if three nested domains are defined.
Each file stores a $n_x\times n_y\times n_t$ tensor for each atmospheric field (such as temperature and precipitation) and vertical level if applicable, where $n_x\times n_y$ is the grid and $n_t$ the temporal discretization. The entries of the matrix are the simulated values in each grid cell. 

\myparagraph{Parameterizations and ensembles.}
Grid resolutions typically used in NWMs cannot capture micro-scale physical processes essential to increase weather simulation quality. %, but the model must consider their effects to simulate atmospheric states. 
For this reason, scientists have created physical parameterizations, \textit{i.e.,} statistical methods, and algorithms aimed at mimicking these effects in NWMs. 
Since the parameterizations are approximations of real physical processes, several approaches are proposed in the literature to represent each physical process. The choice of parameterizations can drastically affect its result. 
Usually, the choice of parameterizations relies on the expert's experience.
In fact, the selection of parameterizations, together with the spatiotemporal discretization and ICBC data generation methodology, are examples of sources of uncertainty in an NWM.
A common approach adopted by experts to reduce their impact is constructing simulation ensembles by fixing the geographical region and the temporal horizon but varying other configuration parameters, such as the spatiotemporal discretization, the ICBC data, and the physical parameterizations~\cite{rautenhaus2018}. 
Although the outputs of a single simulation are deterministic, building an ensemble of simulations enables probabilistic analytical approaches. Also, building simulation ensembles enables the comparison of different results, which is essential in operational weather forecasting~\cite{souza2022}.

\section{Challenges} 
\label{sec:challenges}
Working with numerical weather ensembles, from setting up a single run to analyzing multiple simulations in an integrated way, is challenging. With WRF, starting a run is accomplished by directly editing various text files that store several parameters, opening up many setup possibilities. Configuring a simulation run may become problematic even when most parameters are set to default values. 

First, to delimit nested simulation domains, the user must set non-trivial parameters such as the size, position, and number of cells in each grid used by the model. WRF may fail to find a valid atmospheric state solution if the parameters are not precisely defined. Usually, weather professionals set these parameters through a trial-and-error approach or using tools such as the WRF Domain Wizard. Although helpful, the WRF Domain Wizard can be difficult to use since it is designed to be a general-purpose WRF grid configuration tool.
Second, several parameters must be defined multiple times (in different text files), which is exhaustive and error prone. Although it is important to make WRF a general model, in most real-world scenarios the excess of degrees of freedom is under-utilized. 
A viable workaround to these difficulties is using scripts to automatize the WRF setup phases: edit the configuration files, download ICBC data, and execute the WPS and PRC modules. Systematizing these processes also avoids manually running command line tools, which may be challenging for weather experts. 

After starting a simulation run, experts must wait until its execution is complete to analyze the produced results. However, a single run can take hours or days to complete, depending on the simulation parameters and the available computational environment. Analyzing results \textit{in-situ} could save time, especially for computing-intensive simulations. Nevertheless, it is not recommended to manually access intermediate simulation files during a run since it can interrupt the model and force it to start over. 
Also, a single run generates a large amount of multivariate spatiotemporal data outputs that experts usually analyze using standard tools such as GrADS~\cite{grads}, NCL~\cite{ucar2019}, or Vapor~\cite{li2019}. These tools allow the visualization of atmospheric variables in two-dimensional maps saved as static images. 
%Some software provide three-dimensional and interactive results, even though 3D designs are usually neglected by weather experts~\cite{rautenhaus2018}. 

It is essential to notice that the workflow associated with a simulation is both computing- and data-intensive. Although some popular WRF analysis tools offer dynamic aggregation and interactive data exploration, they are limited to small-scale analyses. In this way, the user may still need to perform laborious tasks to identify patterns and extract information from simulation outputs, made worse when analyzing ensembles. 
It is common for users to run a series of simulations, build ensembles, and evaluate them. However, their construction must be careful since not all simulations have promising results, and bad runs may camouflage valuable information. 
To get around this issue, the user must run and evaluate one simulation at a time before adding them to an ensemble. This task becomes more arduous as the number of simulations increases. 
However, simulations commonly share parameters, and previous configurations may be used as a base for new ones.
Also, waiting for multiple runs to finish and then analyzing the results may take a long time. During the analyses of the results, it is hard to identify patterns while being aware of parameter choices and the uncertainties they impose.
In summary, creating and investigating simulation ensembles pose several challenges. Therefore, research aimed at facilitating these processes is relevant.
We intend to simplify and optimize simulation ensembles' building, management, and analysis in runtime, employing provenance and visualization techniques.

\section{Requirements}
\label{sec:require}
In this work, we have engaged \textcolor{black}{two weather experts} with extensive forecasting experience. 
Through a series of meetings, we discussed with them the requirements a system should meet to make the use of the WRF models simpler and faster.
Our main goal was to simplify their workflow, from building a single run to analyzing the results of simulation ensembles, without restricting their analysis capabilities.
In this sense, we adopted some scope boundaries, which may be expanded in future work to adapt \system{} to other uses. We decided to consider limited-area simulation ensembles where members may differ by the chosen physical parameterizations (what is called physical ensembles) and/or by the source of ICBC data. These two setup possibilities are highly relevant since:
(1)~different parameterizations could lead to highly distinct predictions; 
(2)~studying parameterizations is essential for operational weather forecasting since it requires a deep understanding of their impacts, especially in dealing with extreme weather events;
(3)~initial and boundary conditions that faithfully represent the atmosphere state are crucial. Different ICBC data are rarely identical, and small perturbations may drastically impact predictions.
More precisely, \system{} must satisfy the requirements:

\myparagraph{[R1]~Support simulation setup.}~Enable easy definition of nested domains, start/end dates, ICBC, and parameterization schemes. Since many physical processes exist, our collaborator experts selected the most important for their work: cloud microphysics, cumulus, land surface, surface layer, and planetary boundary layer~(PBL).

\myparagraph{[R2]~Automate simulation runs.}~Manage inputs and outputs of the WRF workflow. The system should automatically start a simulation and allow users to stop or restart runs easily. In addition, if a distributed environment is available (\textit{e.g.}, a small cluster or a computing cloud), the workflow runs can be scheduled to different machines.

\myparagraph{[R3]~Manage provenance data.}~Capture and store the data derivation path of each piece of data, as well as any other metadata associated with each workflow run (the user who started the run, the analysis project it belongs to, the execution time of each step of the workflow, the directory paths used to store the input and output files, \textit{etc.}) Such metadata allows users to track the history of each run.

\myparagraph{[R4]~Optimize preprocessing.}~Automatically decide when there is no need to execute a WRF step and use the output of previous runs (\textit{i.e.}, caching). We assume that the result of an activity of the workflow is stored correctly to be reused if and only if the activity consumes the same parameter values as the previously executed one. For example, suppose an expert wants to perform a new simulation using the same domain configuration as a previous run but considering a different period. In that case, they should not need to rerun the Geogrid step.

\myparagraph{[R5]~Automate the extraction and storage of atmospheric fields.}~Filter, organize and store relevant atmospheric field data from the output files of an in-progress simulation without interrupting the run or damaging the files. The extracted values should be stored in the same database as provenance data, allowing for an integrated analysis.
WRF provides hundreds of output variables. The important ones depend on each application. Our collaborators selected 7 atmospheric fields related to the forecast of extreme rainfall events: precipitation, temperature 2 meters above the surface, divergence at the vertical level of 300~hPa ($\sim$ 10~km above sea level), upward vertical wind at 500~hPa ($\sim$ 5.5~km above sea level), convergence at 850~hPa ($\sim$ 1.5~km above sea level), k-index (an indicator of atmospheric instability), and relative humidity at 850~hPa.

\myparagraph{[R6]~Support the visual analysis of ongoing simulations.}~Allow the interactive visual investigation of the results of in-progress simulations. This requirement would allow the expert to detect patterns early and decide whether it is worth keeping or stopping a run.

\myparagraph{[R7]~Support the creation of ensembles.}~Allow users to seamlessly create as many simulations as necessary and organize them into ensembles according to their technical necessities. For example, if several simulations use the same domain configuration but different ICBC sources, the user should be able to choose to create one ensemble with all runs or multiple ensembles, one for each ICBC source.

\myparagraph{[R8]~Support the visual analysis of simulation ensembles.}~ Visually examine several atmospheric fields in space and time to identify regions and periods associated with relevant patterns and/or target weather scenarios.
\section{\system{} system}
\label{sec:system}
\begin{figure}[!ht]
\centering
\includegraphics[width=\linewidth]{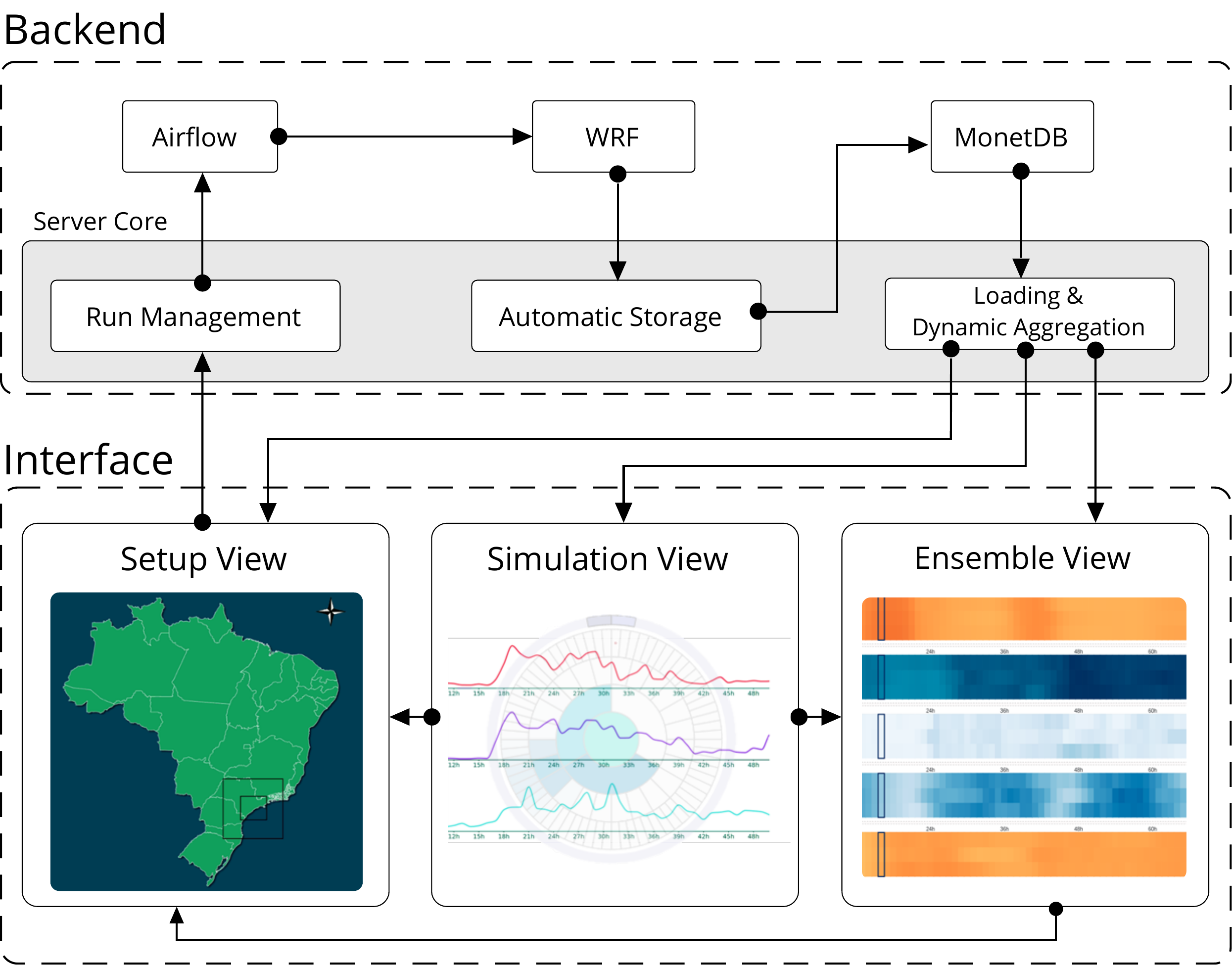}
\caption{Overview of the \system components.}
\vspace{-0.5cm}
\label{fig:sys-overview}
\end{figure}

\system is a web-based visualization system designed to meet the requirements defined by our collaborators. Using a visual and user-friendly approach, the system enables the creation and exploration of weather simulation ensembles based on the WRF model. 
The visual interface includes a set of interactive visual structures to configure~(\textbf{R1}), run~(\textbf{R2}, \textbf{R3}), and analyze~(\textbf{R5}) individual simulations. Also, it provides visual elements to facilitate the construction (\textbf{R6}) and analysis of ensembles~(\textbf{R7}). 
The system also has a backend containing a provenance and metadata database and a workflow management system. The Server Core automatically extracts and stores atmospheric field data~(\textbf{R4}) and manages metadata, input/output workflow files, and WRF runs~(\textbf{R8}).
Using \system, the user can configure and investigate weather simulations by simply using a web browser, without complex command line applications.
\autoref{fig:sys-overview} shows the system's components.

\begin{figure}[!b]
\centering
\includegraphics[width=.85\linewidth]{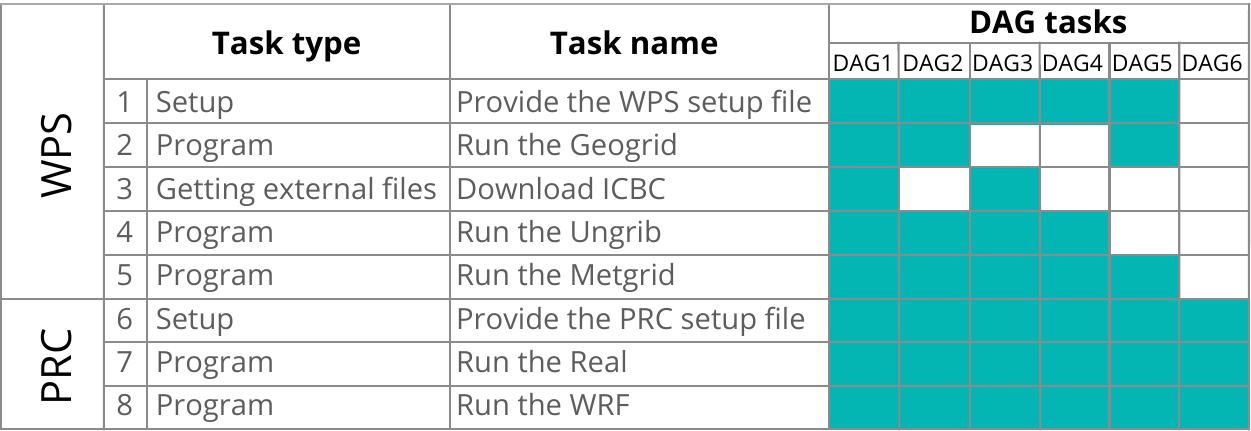}
\caption{\system's DAGs and tasks.}
\label{fig:dag}
\end{figure}

%%%%%%%%%%%%%%%%%%%%%%%%%
%%%     BACKEND
%%%%%%%%%%%%%%%%%%%%%%%%%
\begin{figure*}[!ht]
\centering
\includegraphics[width=\linewidth]{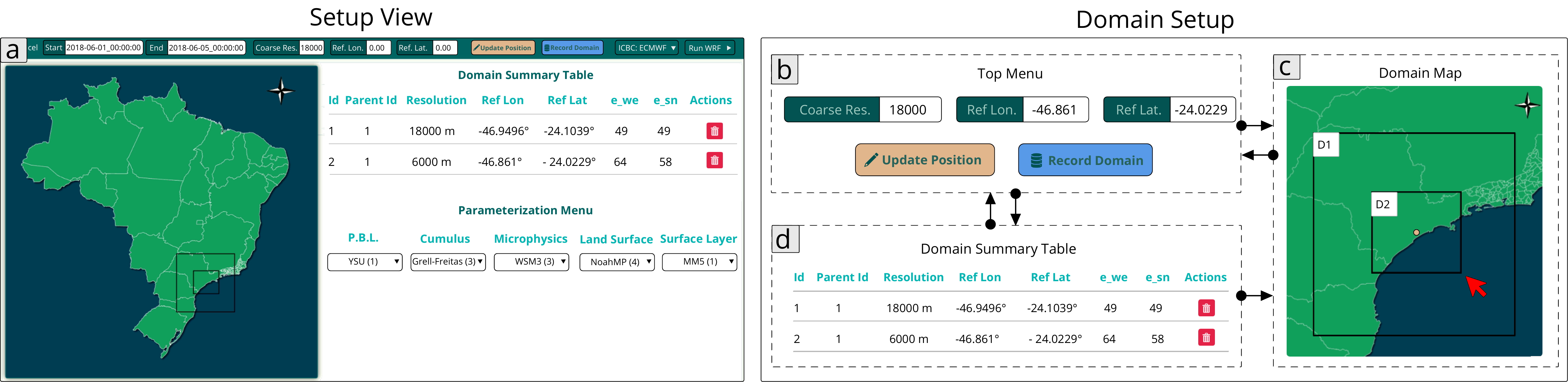}
\caption{(a) The Setup view lets users configure WRF simulations interactively. \textcolor{black}{(b--d)}~Interactions used to set up domains.}
\vspace{-0.4cm}
\label{fig:setupView}
\end{figure*}
\subsection{Backend}
\label{sec:backend}
The backend dynamically manages the WPS and PRC sub-workflows. It records, stores, and loads relevant atmospheric field results, and computes statistical and probabilistic aggregations.
It contains the provenance and metadata database deployed on the column-oriented MonetDB database system. MonetDB was chosen because it performs well with the query types required by \system (\textit{i.e.} it took on average less than 1s to handle queries from the interface in our case studies). It also offers advanced features such as clustering, data partitioning, and distributed query processing that can be explored. The backend also has an embedded workflow system (Apache Airflow) to manage the execution of the WRF workflow. \textcolor{black}{
The Server Core directly connects to the interface using WebSockets, \textit{i.e.}, it can notify the interface when new data is processed, updating the visualizations. The Server Core was built using Python, primarily relying on the libraries Flask-SocketIO, Pymonetdb, NetCDF4, and WRF-Python.}

\myparagraph{Provenance and metadata database}. The provenance and metadata database organizes data in a structured and queryable way. It allows storing information regarding several projects, which can be associated with multiple simulations (\textbf{R3}). The parameters and input/output data are registered in each simulation. Similarly, the metadata associated with the execution of each workflow program (start/end time, errors, \textit{etc.}) is stored. This database is a rich source of information that can be used to optimize the setup of simulations or to provide analytical capabilities to the experts.
Using a database also has made data loading faster since it can handle aggregation or filter queries directly. This strategy is more efficient than loading the entire simulation output to memory and extracting the relevant information afterwards.

\myparagraph{Automatic storage.}\textcolor{black}{
The Sever Core module checks for new files produced by the WRF simulations every minute. When new output files are detected, it stores aggregated information, together with the files' content on MonetDB~(\textbf{R5}), enabling the analysis of simulation results at runtime~(\textbf{R6}).  
The Server Core computes aggregations, \textit{i.e.}, spatiotemporal statistics (minimum, maximum, and average), considering the entire domain and predefined intervals (1~hour, 3~hours, 24~hours, and the complete simulation period). 
Data are stored (and can be retrieved) with information regarding the nested grid, atmospheric field, grid point, and time. 
}

\myparagraph{Dynamic data aggregation.}\textcolor{black}{\system{} handles dynamic aggregation queries.
For example, assume that a user has created an ensemble with five runs and selected three atmospheric fields to analyze in the spatial dimension in a custom period. In this case, given an aggregation function (minimum, maximum, average, and probability), the Server Core computes, at each grid point, the aggregation of each field considering the five runs. If the user modifies the ensemble (\textit{e.g.} adding a member), the backend dynamically updates the aggregations.
}

\myparagraph{Run management.} We adopted the Apache Airflow workflow management system to orchestrate the multiple WRF runs. It allows users to model their workflows as directed acyclic graphs (DAGs), each gathering and organizing tasks to be executed sequentially and/or in parallel (\textbf{R2}, \textbf{R4}).
Our collaborators' daily tasks involve the following: providing the WPS setup file, running Geogrid, downloading ICBC data, running Ungrib, running Metgrid, providing the PRC setup file, and running the Real and WRF programs. Considering all possible workflows that can be modeled based on the combinations of programs, we identified and modeled six DAGs, illustrated in~\autoref{fig:dag}, that are available for execution within \system{}. The first five tasks in the figure are related to the WPS sub-workflow, while the \textcolor{black}{last three} are associated with PRC. 
\textcolor{black}{
\system{} automatically defines what DAG must be used based on the configurations of the current and previous runs:
DAG1 is triggered to generate a new simulation from scratch. DAG2 is triggered when ICBC is available, but the other tasks must be performed. DAG3 is executed when domain settings and intermediate files can be reused from a previous run. DAG4 works under the same assumptions as DAG3, but the ICBC must be available. DAG5 can only be executed when Ungrib outputs can be reused from a previous run. Finally, DAG6 is used when no WPS program needs to be run, \textit{e.g.} the user has only changed the physical parameterizations of a previous run. 
}

%%%%%%%%%%%%%%%%%%%%%%%%%
%%%     FRONTEND
%%%%%%%%%%%%%%%%%%%%%%%%%
% \vspace{-0.2cm}
\subsection{Visual exploration interface}
\label{sec:visualinterface}
\system' interface design and functionalities aim to meet weather experts' technical demands while adopting a user-friendly and visual approach.
Most of its visual components rely on two-dimensional designs that professionals are familiar with, such as scatter plots, line graphs, and heatmaps.
Nevertheless, we also incorporated visual representations that are not so usual to them, such as heat matrices and sunburst charts, since they provide expressive and concise visualizations of simulation outputs.
As shown in~\autoref{fig:sys-overview}, the system interface comprises three interactive views: the Setup, the Simulation, and the Ensemble views. 
When a new analysis project is started, the user is directed to the Setup view to configure and run their first simulation. If a previously created project is loaded, the user can explore previous runs individually or as ensembles of simulations using the Simulation and Ensemble views.

\subsubsection{Setup view}
The Setup view, illustrated in~\autoref{fig:setupView}~(a), was designed to facilitate the setup of a new WRF simulation from scratch~(\textbf{R1}). It comprises a top menu, a domain map, a domain summary table, and a parameterization menu. 
Users can set the simulation domains' grid resolutions and central points using the top menu, define the start/end simulation time, and select the ICBC data source. The domain map and summary table allow the visual definition and verification of the domains' configuration. Finally, the parameterization menu allows the selection of the physical parameterizations considered in the simulation.

The user must follow a simple and visual process to configure the simulation. Once the coarse grid resolution is set in the top menu (\autoref{fig:setupView}~(b)), the user interactively defines a new domain by brushing on the map (\autoref{fig:setupView}~(c)). The latitude and longitude of the grid's central point are linked to the top menu, allowing the user to fine-tune the domain location. We notice that, by using the interface, the users don't need to manually make the parameters of all nested grids compatible since the system automates these tasks.
When a new domain is defined, an overview of its settings (the parent domain's identifier, resolution, latitude and longitude of the central point, and the number of points) is shown in the domain summary table (\autoref{fig:setupView}~(d)). The user can delete a previously created domain through the summary table by clicking the trash icon.
After completing the setup, the user can run the simulation through the top menu. As described in Section~\ref{sec:backend}, the execution and management of the running simulation are automated by the backend~(\textbf{R2}). After the run starts, the Simulation view is loaded.

\subsubsection{Simulation view}
\begin{figure*}[!ht]
\centering
\includegraphics[width=\linewidth]{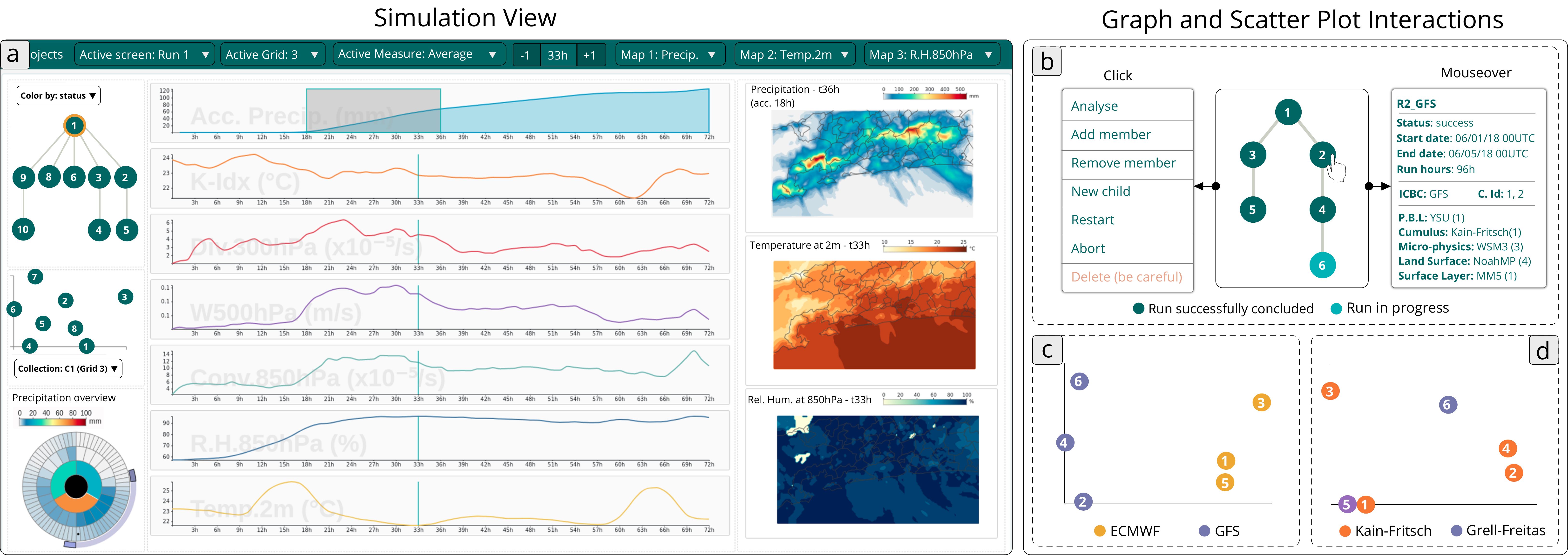}
\caption{(a)~Simulation view. (b)~Example of the runs overview graph (middle) from the São Paulo case study (Section~\ref{subsec:case2}). It shows six runs, 5 completed and 1 in progress. By hovering node 2, the run's information is shown. The user can interact with a run by clicking on its node. (c)~Scatter plot from the same case study colored according to the ICBC source. Runs with the same ICBC generated similar precipitation results; (d)~Scatter plot of the Maricá case study (Section~\ref{subsec:case1}) colored by the cumulus physical process and showing no precipitation pattern.}
\vspace{-0.3cm}
\label{fig:simView}
\end{figure*}

The Simulation view, shown in~\autoref{fig:simView}~(a), provides individual simulation visual analysis capabilities. Its design went through several refinements based on our interactions with weather experts. Its final composition contains a top menu and three panels: the runs overview, the temporal, and the spatial analysis panels. 
The top menu allows the user to configure the currently selected run, the grid considered by the visualizations, the function (average, maximum value) used to compute atmospheric field aggregations over time, and up to three atmospheric variables of interest to be jointly analyzed in the spatial dimension.

The runs overview panel was designed to enable the monitoring of active simulations and the creation of new simulations that share configuration parameters with previous runs. It is composed of a graph and a scatter plot. The graph represents the parent-children relationship among runs. Each node of the graph represents a different simulation run. The graph's root is constructed using the Setup view. The setup data of any node can be reused to create a child node representing a simulation that shares setup parameters with its parent~(\textbf{R4}).
For example, in~\autoref{fig:simView}~(b), runs 2 and 3 were based on run 1. Similarly, run 3's parameters were the basis for the construction of run 5; run 2 was used as the starting point of the setup of run 4, \textit{etc.} The idea is that the structure of the provenance database mirrors the structure of this graph. 
Each node of the graph has mouse click and mouse over implementations. As shown in~\autoref{fig:simView}~(b), when the mouse is over a node, its primary information (name, status, simulation start/end date, ICBC metadata, physical parameterizations, \textit{etc.}) is displayed.
If a node is clicked, the user can choose to \textit{Analyze} the simulation in the Simulation view, even when it is incomplete~(\textbf{R6}); create a \textit{New child} reusing its setup parameters~(\textbf{R1}, \textbf{R4}); \textit{Restart} or \textit{Abort} a simulation that presented failures or has unpromising results~(\textbf{R4}); and \textit{Delete} a simulation and its graph node~(\textbf{R7}). It is also possible to \textit{Add} or \textit{Remove} the simulation from an ensemble of simulations~(\textbf{R7}).

The scatter plot, illustrated in \autoref{fig:simView}~(c) and~(d), shows how similar the performed simulations are considering the predicted precipitation volume. Each circle in the chart represents a simulation; the closer the dots, the more similar precipitation volumes they have. 
\textcolor{black}{To construct the scatter plot, we computed feature vectors representing each simulation. The vectors contain the statistics of the predicted rainfall volumes (maximum, average, and standard deviation) considering time intervals of 3h, 24h, and the entire simulation period. So, each simulation is represented by a 9-dimensional feature vector.
Then, we applied the principal component analysis (PCA) method to project the feature vectors. We decided to use PCA since the experts were already familiar with the method, but any other projection technique could be used.}
This visualization helps the user relate the precipitation simulation output to its setup parameters. This relation is important since the user can make sense of the parameters' sensitivity and use this knowledge as a guide to create simulation ensembles~(\textbf{R7}).
The user can color the graph nodes and scatter plot dots by the run's status (\textit{i.e.} success, running, or failed -- \autoref{fig:simView}~(b)), the parameterizations (\autoref{fig:simView}~(c)), the ICBC data source (\autoref{fig:simView}~(d)), and the containing ensemble's id. 

\begin{figure}[b!]
\centering
\includegraphics[width=\linewidth]{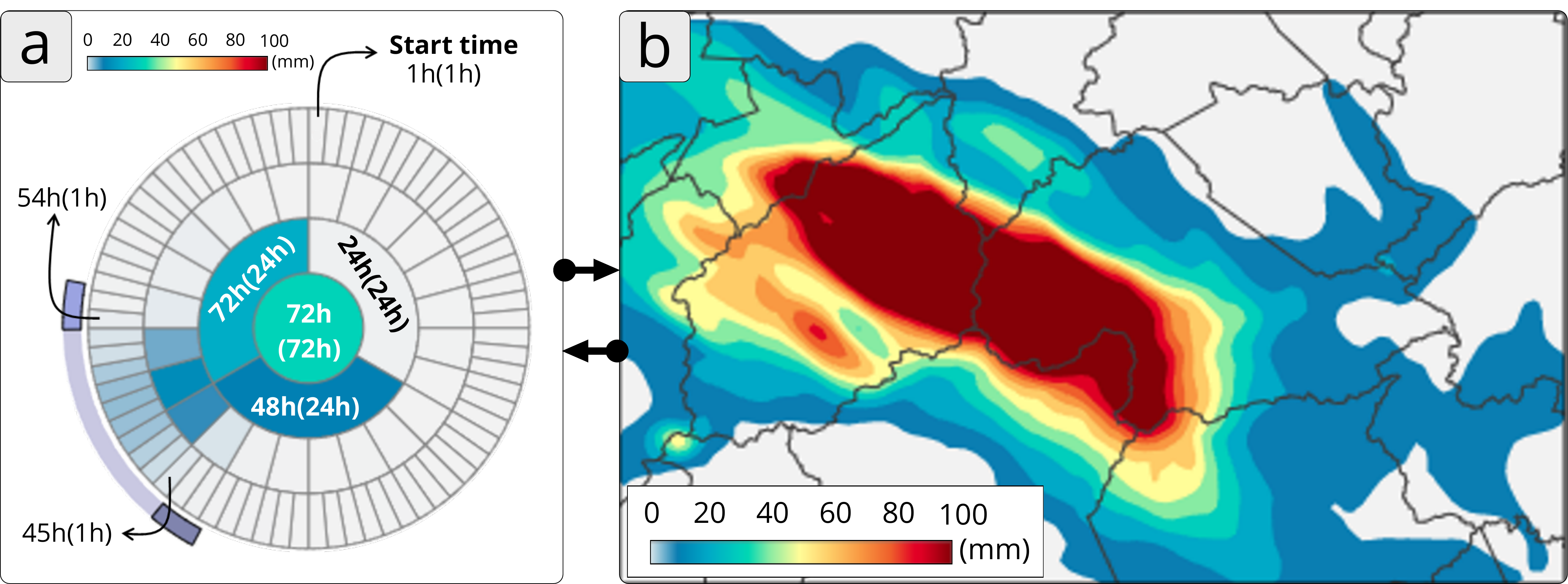}
\caption{The map updates if a time interval is set in the sunburst chart. Conversely, the sunburst chart is recomputed if a grid point is clicked.}
\label{fig:sb}
\end{figure}

The Simulation view implements the temporal analysis panel to enable the time-based analysis of simulations. The panel is composed of a collection of line charts (middle column in~\autoref{fig:simView}~(a)) and a sunburst chart (bottom left chart in~\autoref{fig:simView}~(a)).
The line charts show the simulated value of different atmospheric fields at each time step. Weather experts widely use this type of visual representation, especially to study accumulated rainfall, because it allows them to quickly identify when rainfall events may happen and their magnitude.
The sunburst chart is a collection of concentric circles representing different time aggregations and is colored based on the accumulated precipitation in each period.
The outer cells represent one-hour intervals. The second group of cells represents three-hour intervals. Following this, the third layer represents 24-hour intervals, and the central cell represents the entire time horizon. Although weather experts are not familiar with this visualization, the extra information about accumulated values is fundamental to interpreting and identifying high-volume precipitation events. 
In~\autoref{fig:sb}~(a), the cell identified as \textit{72h (24h)} represents the rainfall accumulated between the 48h and the 72h simulation steps, \textit{i.e.} the rainfall accumulated in the 24 hours immediately before the 72h time step. 
In the example, the time horizon has 72 hours. Because of that, the sunburst has 72 one-hour cells in the first layer, eight 3-hour cells in the second layer, three 24-hour cells in the third layer, and one 72-hour cell in the center. The user can interact with the chart by clicking and brushing the cells. Doing so, the spatial rainfall distribution shown in the spatial analysis panel is updated to reflect the selected period (see~\autoref{fig:sb}~(b)).

Finally, the spatial analysis panel allows the analysis of the distribution of atmospheric fields. It contains three maps for the visualization of the distributions of different fields at a given instant (or the accumulated field distribution over a time interval, in the case of precipitation volumes). 
\textcolor{black}{We use the marching squares algorithm to generate the contours, primarily because of its efficiency.}
The user can control the time reference used to build the maps using the top menu and, in the case of the rainfall field, using the sunburst.
The active period is indicated in the line charts by a vertical line. In the case of accumulated rain volumes over a time range, a rectangle is used to indicate the selected time interval.
If the user wants to investigate a specific domain point, they can click on a map to select it. In doing so, the temporal analysis panel is updated only to show data associated with the selection.

\subsubsection{Ensemble view}
\begin{figure*}[!ht]
\centering
\includegraphics[width=\linewidth]{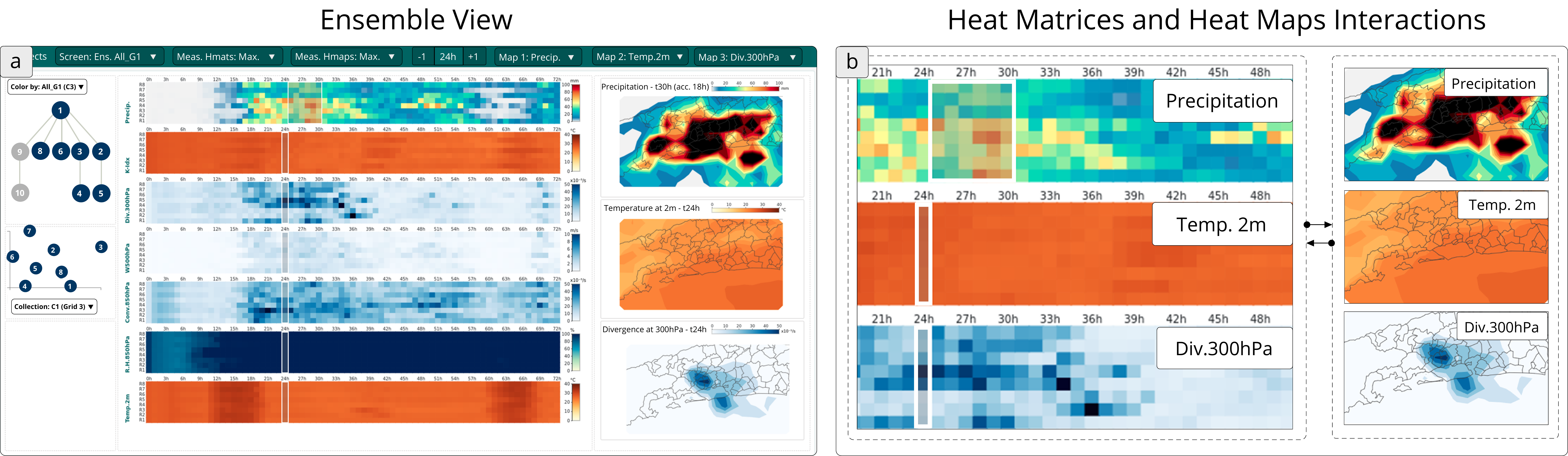}
\caption{(a)~Ensemble view. (b)~Examples of precipitation, temperature at 2m, and divergence at 300~hPa heat matrices (from top to bottom). The interval of 25h-30h (precipitation matrix) and step 24h (other matrices) are selected. (c)~Accumulated rainfall, temperature at 2~m, and divergence at 300~hPa spatial distributions.}
\vspace{-0.3cm}
\label{fig:ensView}
\end{figure*}

Ensemble view, illustrated in~\autoref{fig:ensView}~(a), was designed to enable the temporal and spatial visual analysis of dynamically constructed ensembles, regardless of the number of members~(\textbf{R7}, \textbf{R8}). \textcolor{black}{Similar to the Simulation view}, its final design was defined in conjunction with experts. It contains a top menu, a temporal, and a spatial analysis panel.
\textcolor{black}{The Ensemble view uses the same spatial visualizations provided in the Simulation view.
The only difference to the maps in the Simulation view is that it aggregates values of multiple ensemble members.} Experts can explore worst and average scenarios by changing the aggregation function. Using similar metaphors facilitates the experts' interpretation of results and increases their engagement with the system.

The temporal analysis panel comprises a collection of heat matrices, one for each atmospheric field. 
Each row (y-axis) of a matrix represents an ensemble member, while each column (x-axis) is associated with a time step. The color of each cell is mapped to the aggregated value of the simulated atmospheric field over the domain (or a grid point selected using the map). 
Using these matrices was important to enable the visualization of multiple ensemble members simultaneously while still encoding the global ensemble patterns.

The time information used to build the maps is represented in the heat matrices, as illustrated in \autoref{fig:ensView}~(b, c). In the example, the time step 24h (highlighted in the last two matrices) is used to construct the maps associated with the temperature at 2 meters and divergence at 300~hPa variables. 
Using the precipitation volume matrix, the user can brush an interval to study accumulated rainfall values (top of~\autoref{fig:ensView}~(b)). The precipitation map shows the maximum volumes observed between 24h and 36h.
We observe that different time references can be used to analyze rain volumes and other variables. 
This feature is vital since atmospheric fields are interpreted to investigate future precipitation.

In addition to using the statistical aggregation functions, the Ensemble view also allows the computation and visualization of the probability of observing scenarios of interest. For example, the user can use the matrices to visualize the likelihood of observing temperatures higher than a certain threshold. 
To do so, the expert must define the target values for each variable that characterizes the scenario of interest. Each matrix cell displays whether such a scenario occurred at least in one domain point.
When computing rainfall probabilities, the user must define a value threshold and a period over which the precipitation accumulation should be considered (for example, investigate the likelihood of 100~mm occurring in one hour). 
This information is essential since 100~mm of rain in one hour is considered extreme, while 100~mm of rain spread over 7 days is not important. 
The user can also define the period when the event may happen. For example, the expert may be interested in verifying the probability of a condition (\textit{e.g.}, rainfall volume of 100~mm/h) in a specific time window (\textit{e.g.}, the first 24 hours) to check the probability of observing 100~mm/h rainfall in the first 24h of simulation. This analysis may be used to trigger weather alerts.

\vspace{-0.1cm}
\section{Case studies}
\label{sec:case}
\textcolor{black}{Two weather experts} conducted two case studies to validate the system. They used a desktop with a Ryzen 7 3700X 3.6GHZ, 16GB, and GeForce GT 210 1GB.
The first study refers to an extreme rainfall event in Maricá, a city in the state of Rio de Janeiro, Brazil.
The second analyzes a rainfall event in São Paulo, Brazil.

\subsection{Extreme rainfall event in Maricá (2022)}
\label{subsec:case1}

\system was initially used to evaluate WRF forecasts of an extreme rainfall event that happened on April 1, 2022, in Maricá (white pin in~\autoref{fig:teaser}~(d)).
This event was associated with a cold front that moved through the southeastern region of Brazil between March 31 and April~2,~2022. The greatest rain volumes were observed in the western area of Maricá from 10PM on March~31 to 3AM on April~1 and in the city's central region between 12AM on April 1 and 3AM on April 2 (GMT). The weather stations recorded accumulated rainfall volumes between 88mm and 260mm at these moments, accounting for the previous 24 hours. These volumes characterize an extreme storm and several people lost their homes. 

The weather experts started the investigation by creating a new project on \system. Then, using the Setup view, they built their first simulation, in which the main configuration parameters are illustrated in~\autoref{fig:teaser}~(a). Using the top menu, they entered the start (03/31/2022 00:00~GMT) and end dates (04/03/2022 00:00~GMT), the coarser domain's spatial resolution (18,000 meters) and selected GFS as the source of the ICBC data. Afterwards, they defined three nested domains using the domain map. The experts checked the domain summary table for each demarcated domain to ensure that the grid's resolution and the position of their central points were adequately defined. Eventually, they preferred to type the latitude and longitude of the central point using the top menu to adjust its location precisely. Then, they selected one parameterization for each physical process they wanted to consider in the simulation and clicked the \textit{Run WRF} button to start it. They were automatically directed to the Simulation view and observed that the runs overview graph was updated with a new root node.

As the simulation evolved, the experts evaluated the results and decided to keep the simulation running. If the simulation results were not promising, \system could be used to stop the simulation to save computational resources. After analyzing the outputs using the temporal analysis panel (configured to use the maximum aggregation function), the experts identified that some atmospheric variables indicated an extreme event between time steps 18h and 36h. As red boxes highlight in~\autoref{fig:teaser}~(b), the divergence at 300~hPa, vertical upward wind at 500~hPa, convergence at 850~hPa, and the precipitation itself were very high, especially in the finer grid resolution.

To deepen the analysis, the weather experts selected several different time intervals using the sunburst chart (see~\autoref{fig:teaser}~(c)) to inspect the accumulated rain spatial distribution in the spatial analysis panel. During this process, they observed that although the model predicted high rainfall volumes, they were concentrated in other Rio de Janeiro areas. In fact, they noticed that the amount of rain predicted for Maricá was not compatible with the rain volumes observed in the city during the event (see~\autoref{fig:teaser}~(d)).
By selecting points in Maricá on the map, the line charts/sunburst were automatically updated to show the associated data. The experts confirmed that the model could not predict the event's magnitude in the most affected areas throughout the time horizon. 

Despite that, the interpretation of other variables still suggested the possibility of heavy rainfall in the city.
Aiming at better reproducing the event, the weather experts derived child runs from the root node using the runs overview graph. The new runs reused the base configuration of the first simulation but considered different physical parameterizations. A total of eight runs were created. \autoref{fig:teaser}~(e) shows the final graph colored by the cloud micro-physics parameterization choice.
By inspecting the overview scatter plot, no particular pattern was identified on the runs precipitation prediction considering their parameterizations (see~\autoref{fig:teaser}~(f)). Furthermore, very few differences between the runs were identified by individually exploring each run using the Simulation view. As the next step, the experts decided to compose an ensemble.
Using the Ensemble view to evaluate the probability of observing an extreme event in Maricá, the following threshold values were defined: K index of 27°C, precipitation of 40mm accumulated in 1h, divergence at 300~hPa of 30$\times10^{-5}$/s, upward vertical wind of 5.0~m/s, convergence at 850~hPa of 30$\times10^{-5}$/s, relative humidity of 100\%, temperature at 2~m from the surface of 30°C. 
Observing the heat matrices, the experts identified a possibility of rainfall volume of 40mm/h between 18h and 36h (see~\autoref{fig:teaser}~(g)). Observing the spatial analysis panel, they verified that although the possibility was relevant in some areas of the domain, it was very subtle in Maricá (see~\autoref{fig:teaser}~(j)).
Similarly, the simulated K index values were greater than 27°C from 0h to 27h (see~\autoref{fig:teaser}~(h)), but these values were mainly observed outside Maricá (see~\autoref{fig:teaser}~(k)). 
Finally, relative humidity achieved 100\% in some regions at time step 13h (see~\autoref{fig:teaser}~(i)). However, although the experts observed high probabilities of reaching the target value in most areas of the domain, in Maricá, these values could only be observed at time step 30h (see~\autoref{fig:teaser}~(h)).
Although the simulation results could not indicate the possibility of an extreme event in Maricá, the experts could easily run and analyze several simulation scenarios without worrying about setup, data management, and visualization technical details that would be required if the same experiment were performed without \system. False-negative weather predictions are why atmospheric modeling remains an active research field.

\subsection{Rainfall event in São Paulo (2018)}
\label{subsec:case2}

In 2018, a frontal system moved across São Paulo, causing rain. In this case study, the weather experts used \system to test the WRF model's sensitivity to different ICBC, grid resolutions, and physical parameterizations.
The meteorologists performed six runs from 06/01/2018 00:00~GMT to 06/05/2018 00:00~GMT, a 96-hour interval (see~\autoref{fig:simView}~(a)). They set up two domains, the first with a grid of 18~km and a nested one of 6~km (the same procedure described in Section~\ref{subsec:case1}).
The six runs used three different parameterization combinations for the cloud microphysics, cumulus, and PBL physical processes (\textit{i.e.} the first/second, third/fourth, and fifth/sixth runs shared the same parameterizations). Also, the odd-numbered runs used ECMWF as the ICBC data source, while the even-numbered ones used GFS.

By coloring the points of the runs overview scatter plot according to the ICBC source, the experts identified that considering the coarser grid, simulations that generated similar results used the same ICBC data (\autoref{fig:simView}~(c)). In other words, it was possible to observe two clusters, one for each ICBC data source.
Similar patterns were observed in the finer grid (\autoref{fig:simView}~(d)), but interestingly, runs 2 and 4 were closer than in the coarser grid, which shows that the grid resolution significantly affects rainfall predictions.
Moreover, after a closer look at the clusters, the experts noticed that runs 1 and 5 were the closest ones in both cases (\autoref{fig:simView}~(c) and~(d)), indicating that they have similar precipitation outputs. This finding was surprising since these runs use different parameterizations for all physical processes. 
In contrast, runs 3 and 5 used the same parameterizations for PBL and cloud microphysics, and runs 1 and 3 used the same for cumulus. The experts considered this finding interesting because it shows that the results are sensitive not necessarily to the parameterizations individually but to their combination.

\begin{figure}[!ht]
\centering
\includegraphics[width=\linewidth]{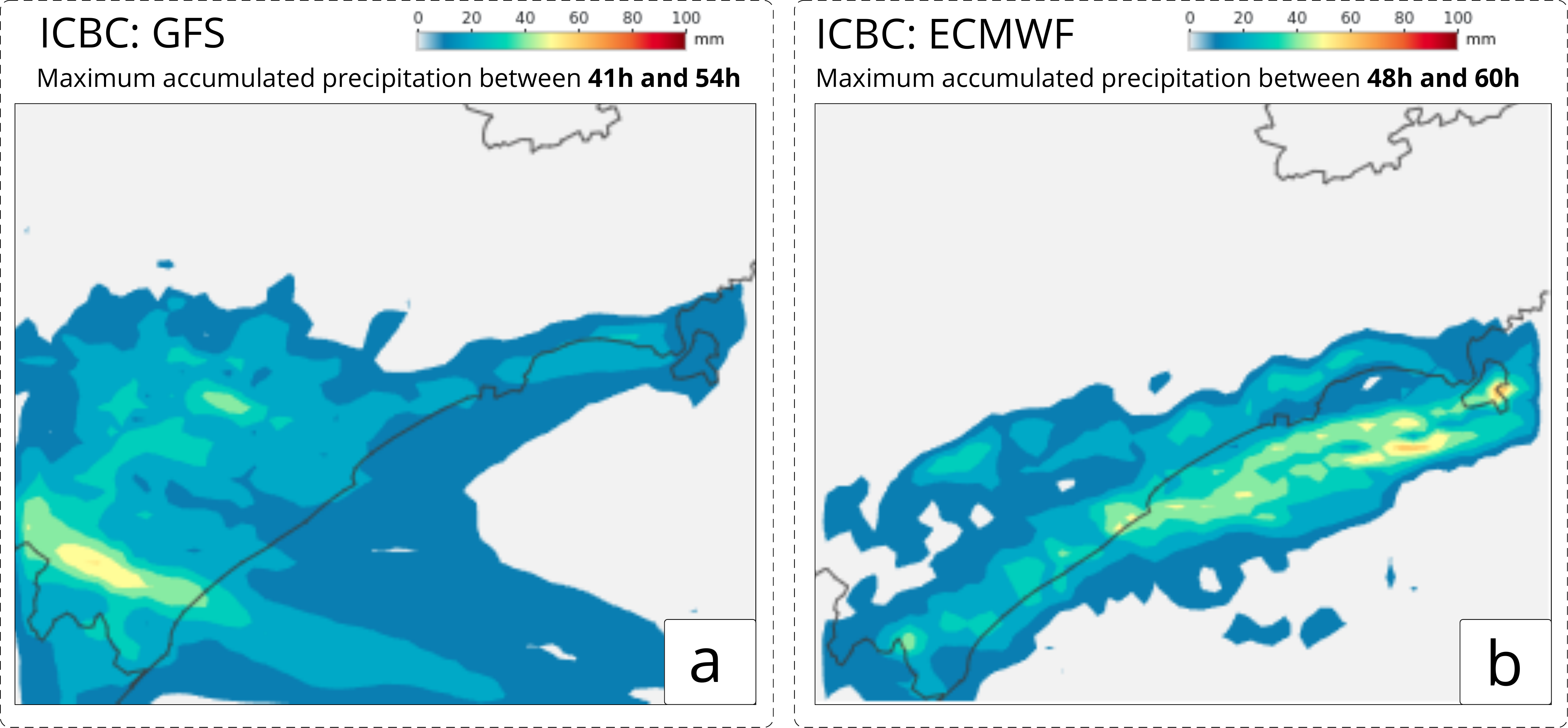}
\caption{São Paulo case study: The maximum accumulated rainfall was between 41h and 54h, according to the ensemble members constructed with the GFS data~(a), and between 48h and 60h, according to the ensemble members who used the ECMWF data~(b).}
\vspace{-0.5cm}
\label{fig:spCase}
\end{figure}

Overall, the analyses of each run in the Simulation view showed that the model predicted an underwhelming event regarding rainfall volume and storm cloud formation.
The experts created three ensembles to strengthen the investigation of the ICBC data influence on the results. Each ensemble was composed of exactly two members configured using the same parameterizations and different ICBC (\textit{i.e.} ensemble 1 contained runs 1 and 2; ensemble 2, runs 2 and 4; and ensemble 3, runs 5 and 6).
The three ensembles showed similar behaviors. In general, ensembles using GFS ICBC data had rainfall forecasts between 41h and 54h on the coast and in the southern part (\autoref{fig:spCase}~(a)). In comparison, the ones using ECMWF data produced rainfall forecasts between 48h and 60h, mainly on the coast of São Paulo (\autoref{fig:spCase}~(b)).

Next, the meteorologists created an ensemble containing all runs. By analyzing the maximum precipitation values, it was possible to visualize (in the spatial analysis panel) a worst-case scenario of heavy rain between 41h to 60h. However, the experts did not consider it an extreme event since the maximum values were up to 70~mm in 20 hours. These predictions would not generate a severe event alert in an operational forecast scenario.
Regarding the other variables, the simulations using the GFS data produced higher values for divergence at 300~hPa, vertical upward wind at 500~hPa, and relative humidity at 850~hPa than the ones using ECMWF data. The results were similar for the other atmospheric fields. Following the analysis of rainfall volumes, the predicted values would not justify the declaration of attention stage by the city authorities. In fact, high values of relative humidity (\textit{e.g.}, 100\%) until time 39h and a temperature at 2~m of 40$^\circ$C at some points of the region indicated the possibility of rain, but not a storm.
Even considering all variables and the ensemble with all runs, there was no relevant probability of extreme event occurrence. Overall, the experts found the strategies for building and analyzing ensembles provided by \system{} very helpful since they could easily compare multiple scenarios and avoid misinterpretations of the simulation results.
\section{Experts' feedback}

According to the experts, \system{} combines attributes that support their work in many ways. First, they said that it provides excellent assistance in setting up a simulation, especially regarding domain delimitation and grid construction, which are complex tasks even for experienced users. 
They also agreed that the approach used for defining physical parameterizations, setting the simulation time horizon, and downloading the ICBC data is more straightforward than editing the WPS and PRC configuration files. Although manually editing files is not intellectually complicated (except for the construction of the grids), it can become a tricky and error-prone task.
Moreover, they reinforced that by automating the step-by-step execution of a run, the system saves experts' time. This is especially true when an error occurs during WPS or PRC execution since the system interface facilitates the identification of configuration errors. They said these capabilities alone make \system{} a huge improvement for their daily workflow.

Another good experience reported was the possibility of visualizing the outputs of an ongoing simulation since it allows the evaluation of the results without waiting for hours (or even days) to complete a simulation. 
Usually, the experts avoid touching incomplete output files to prevent file corruption. When the inspection is required, experts must be extremely careful during the inspection of partial results; otherwise, they can invalidate runs that have already consumed computational resources for an extended period. In this sense, this capability of \system{} represents a substantial contribution in their opinion.
They looked favorably at reusing files and data from previous runs in a controlled, safe, and interactive way. They considered that advantageous due to the time it potentially saves when managing related runs, \textit{e.g.}, those using the same domains, parameterizations, or ICBC data.
They have approved the automatic organization of input/output files and data by users and projects so several experts can use the system simultaneously. In addition, since \system{} keeps the original WRF files, they pointed out that it is possible to use them for other purposes besides the system, not restricting the experts' work.

Regarding the analyses, the experts agreed that the available visualizations and interactions greatly favor the rapid exploration of a simulation in space and time. \system{}' organization and quick response to user requests facilitate the cognitive processing of the simulation results. 
In their opinion, the interface groups familiar visual structures, such as line graphs and heat maps, making the system user friendly. They considered the sunburst chart a novelty and took some time to understand its usefulness. After they became familiar with the visualization, they said it helped to visualize the accumulated rainfall at different intervals. 
Another positive feedback was related to the dynamic creation of ensembles. They said it enriched their ability to explore ensembles with members selected based on different criteria. They also said that the heat matrices were unfamiliar. However, they enjoyed the experience of visualizing ensembles as a whole. The heat matrices coupled with the maps helped evaluate and compare different runs. The experts also appreciated the identification of custom-defined scenarios provided by visualizations. 
\textcolor{black}{Finally, the experts questioned the system overhead, since WRF simulations are already costly.
In fact, computing the simulations dominates the execution time (207 minutes for the first use case and 27 minutes for the second one). The computation by the Server Core is 3 to 5 times faster (72 minutes for the first use case and 6 minutes for the second one).
Since both tasks run in parallel, \system adds no overhead regarding running time.
}

The experts contributed with suggestions for the improvement of \system. Currently, the system allows users to select a grid point on the map, and experts consider the feature essential. However, they would like to be able to brush custom areas of the map. 
Another suggestion is to allow the user to freely define the atmospheric fields of interest. They commented that some professionals are used to inspecting specific variables, and their unavailability may limit the use of the system. We report that this functionality is straightforward to implement. The selection of particular fields was based on our collaborators' needs and was implemented to reduce the scope of our prototype implementation.
In addition, they said it would be even more interesting to import WRF runs that were neither configured nor executed using \system, \textit{i.e.}, use the system to explore runs manually created and previously executed by the experts. This would make the system appealing to a larger audience.

\section{Conclusion and future work}

\system was designed to facilitate the setup, execution, management, inspection, and analysis of WRF runs and dynamically create ensembles, considering different ICBC data, physical parameterizations, and domain configurations. 
The system was constructed as a client-server web application. The backend comprises a MonetDB database, the Apache Airflow workflow system, the WRF model, and a server core. The database stores metadata regarding users, projects, and weather simulations and controls data provenance to enable future queries. The workflow system optimizes the modeling process. The server core connects those modules, extracts, and automatically stores relevant atmospheric fields in the database, and organizes input/output files. Also, it responds to the interface's requests related to single and ensembles of simulations, which usually involve dynamic data aggregations. The system interface consists of three main views: one for setting up a run, one for exploring a simulation, and one for exploring an ensemble.
The Setup view, and the entire process behind it, allows the user to save time and effort during the setup and execution of a simulation. This approach facilitates the development of studies in meteorology because it takes the focus away from the model execution, which is laborious in itself, and allows the user to devote time to the analysis of the generated results.
The Simulation and Ensemble views provide visual structures that help manage the runs, inspect their outputs, and even perform similarity analysis to identify patterns. Both views offer visualizations that allow temporal and spatial aggregations using statistical and probabilistic metrics.

With \system{}, two case studies were performed considering rainfall events caused by cold fronts. During their realization, it was possible to inspect multiple simulation outputs effortlessly, even when the model was running. The experts could use the visualizations to analyze spatiotemporal patterns and compare the results of several simulations. The WRF results did not indicate the possibility of extreme events in the areas of interest. The experts used the system to argue that these simulations provided false negative results. Their analyses reinforce the need for studies to improve atmospheric modeling.
Given the results and experts' feedback, \system{} met its primary purpose: to aid weather analysis through data visualization and provenance.

In future work, we plan to extend the system so it can be used to configure any WRF simulation. We also plan to propose other visualizations and interactions that may take the visual exploration of the simulation ensemble a step further. Future versions of \system can also provide a specialized workflow scheduler component to execute the workflows in parallel. This mechanism can benefit from heterogeneous environments to speed up workflow execution. Also, the provenance data can be used for recommending \system configurations for novice users based on the previous runs configured by weather experts.

\section*{Acknowledgments}
We would like to thank the reviewers for their constructive comments and feedback. This study was partly funded by CNPq (316963/2021-6), FAPERJ (E-26/202.915/2019, E-26/211.134/2019), CAPES (Finance Code 001), and the University of Illinois' Discovery Partners Institute.
\vspace{-0.6cm}

\bibliographystyle{abbrv-doi-hyperref}

\textcolor{black}{
\bibliography{references}
}
% %%%%%%%%%%%%%%%%%%%%%%%%%%%%%%%%%%%%%%%%%%%%%%

\end{document}